\documentclass[aps,pra,floatfix,showpacs,twocolumn,superscriptaddress]{revtex4}

\usepackage{graphicx}
\usepackage{latexsym}

\newcommand{\mypsi}{\psi}
\newcommand{\myE}{\mathbf{E}}

\begin{document}

\title{Quantum states far from the energy eigenstates of any local
  Hamiltonian}

\author{Henry L. Haselgrove}
\email{hlh@physics.uq.edu.au}
\affiliation{School of Physical Sciences, The University of Queensland,
Queensland 4072, Australia}
\affiliation{Institute for Quantum Information, California Institute
of Technology, Pasadena CA 91125, USA}
\affiliation{Information Sciences Laboratory, Defence Science and
Technology Organisation, Edinburgh 5111, Australia}
\author{Michael A. Nielsen}
\email{nielsen@physics.uq.edu.au}
\homepage[\\ URL:]{http://www.qinfo.org/people/nielsen/}
\affiliation{School of Physical Sciences, The University of Queensland,
Queensland 4072, Australia}
\affiliation{Institute for Quantum Information, California Institute
of Technology, Pasadena CA 91125, USA}
\author{Tobias J. Osborne}
\email{T.J.Osborne@bristol.ac.uk}
\affiliation{School of Physical Sciences, The University of Queensland,
Queensland 4072, Australia}
\affiliation{School of Mathematics, University of Bristol,
University Walk, Bristol BS8 1TW, United Kingdom}

\date{\today}

\begin{abstract}
  What quantum states are possible energy eigenstates of a many-body
  Hamiltonian?  Suppose the Hamiltonian is \emph{non-trivial}, i.e.,
  not a multiple of the identity, and $L$-\emph{local}, in the sense
  of containing interaction terms involving at most $L$ bodies, for
  some fixed $L$.  We construct quantum states $\mypsi$ which are
  ``far away'' from all the eigenstates $\myE$ of any non-trivial
  $L$-local Hamiltonian, in the sense that $\| \mypsi - \myE \|$ is
  greater than some constant lower bound, independent of the form of
  the Hamiltonian.
\end{abstract}


\maketitle

%
%
A central problem in physics is the characterization of eigenstates of
many-body Hamiltonians.  Less attention has been devoted to the
complementary question: which quantum states are not the eigenstates
of any physically plausible Hamiltonian?  The purpose of this paper is
to address this question, by explicitly constructing states which are,
in a sense made precise below, far away from the eigenstates of any
non-trivial, local Hamiltonian.  Such constructions are interesting
for several reasons.  First, they place fundamental restrictions on
the physics of many-body quantum systems.  Second, as we discuss in
detail below, our construction gives insights into the construction of
``naturally fault-tolerant'' quantum systems that are able to resist
the effects of noise and decoherence.

%
%
The paper begins with a simple counting argument showing that ``most''
quantum states are not the eigenstates of any physical Hamiltonian.
We then give a more powerful --- albeit, still quite simple ---
argument constructing quantum states $\mypsi$ ``far away'' from all
the eigenstates $\myE$ of any non-trivial, $L$-local Hamiltonian.  In
this statement, by non-trivial we mean not a multiple of the
identity~\footnote{Obviously, all quantum states are eigenstates of a
  Hamiltonian which is a multiple of the identity.}, and by $L$-local
we mean that each interaction term in the Hamiltonian involves at most
$L$ bodies.  Of course, physically we expect that $L$ is a small
constant, $2$, or at most $3$ in special circumstances.
Quantitatively, for an $n$-body system whose constituents have
$d$-dimensional state spaces, we prove $\| \mypsi - \myE \| \geq
\left[ (L+1){ n \choose L} (d^2-1)^{L} \right]^{-1/2}$.  What is
interesting about this bound is that it is a \emph{constant} lower
bound that holds for the eigenstates of \emph{all} non-trivial
$L$-local Hamiltonians, even those with degenerate eigenstates.  

%
%
It is worth noting that the results reported in this paper hold
unchanged for any $n$-local \emph{observable}, not just Hamiltonians.
However, particularly in the light of recent work characterizing the
entangled properties of the ground states of lattice
systems~\footnote{See, for example,
  \cite{Osborne02a,Osterloh02a,Vidal02b,Gunlycke01a}, and references
  therein.}, the case of the Hamiltonian is of especial interest, and
we prefer this nomenclature throughout.

%
%
Interestingly, the states $\mypsi$ we construct are special examples
of \emph{quantum error-correcting
  codes}~\footnote{See~\cite{Nielsen00a,Preskill98c} for a review and
  references.}; such codes turn out to be rich sources of states which
are not close to being eigenstates of any non-trivial, local
Hamiltonian.  Our paper thus illustrates a general idea discussed
elsewhere~\cite{Osborne02b,Nielsen02e,Preskill00a,Nielsen98d}, namely,
that quantum information science may provide useful tools and
perspectives for understanding the properties of complex quantum
systems, complementary to the existing tools used in quantum many-body
physics.

%
%
We begin with a counting argument showing most quantum states cannot
arise as energy eigenstates of local Hamiltonians.  This counting
argument has the advantage of simplicity, but also has some
significant deficiencies, discussed and remedied below.  Suppose an
$n$-body quantum system is described by an $L$-local Hamiltonian, $H$.
We suppose, for simplicity, that each quantum system has a
$2$-dimensional state space, that is, the systems are ``qubits'', in
the language of quantum information science.  It is straightforward to
adapt the argument below when the component systems have state spaces
with higher dimensionalities, and also when different systems have
different dimensionalities.

%
%
It will be convenient to expand our Hamiltonian as
\begin{eqnarray} \label{eq:op-expansion}
  H = \sum_\sigma h_\sigma \sigma,
\end{eqnarray}
where $h_\sigma$ are real coefficients, and the $\sigma$ denote tensor
products of the Pauli matrices $I, \sigma_x, \sigma_y, \sigma_z$.  For
an $L$-local Hamiltonian, we see that $h_\sigma = 0$ whenever the
\emph{weight} of $\sigma$ --- that is, the number of non-identity
terms in the tensor product --- is greater than $L$.

%
%
The number of independent real parameters $h_{\sigma}~$\footnote{For
  systems of dimension $d > 2$ the parameters $h_{\sigma}$ may
  be complex, depending on the operator basis used in
  Eq.~(\ref{eq:op-expansion}).  However, a similar argument to that
  below shows that the number of independent real parameters is still
  given by Eq.~(\ref{eq:number}), but with $3$ replaced by $d^2-1$.}
occurring in Eq.~(\ref{eq:op-expansion}) is:
\begin{eqnarray} \label{eq:number}
  \#(n,L) = \sum_{j=0}^L {n \choose j} 3^j. 
\end{eqnarray}
To see this, note that the different terms in the sum come from the
interactions involving $j=0,1,\ldots,L$ bodies, respectively.  For the
$j$-body interactions, there are ${n \choose j}$ ways of picking out a
subset of $j$ interacting systems, and given a particular subset the
number of parameters is $3^j$, corresponding to the $3^j$ non-trivial
tensor products of Pauli operators.  When $L \leq n/2$ we obtain a
useful upper bound on $\#(n,L)$ by noting that ${n \choose j} \leq {n
  \choose L}$, and $3^j \leq 3^L$:
\begin{eqnarray} \label{eq:number-params}
  \#(n,L) \leq (L+1) {n \choose L} 3^L. 
\end{eqnarray}
For real physical systems we expect $L = 2$ or (rarely) $L = 3$, for which:
\begin{eqnarray}
  \#(n,2) & = & \frac{9n^2-3n+2}{2}; \\
  \#(n,3) & = & \frac{9n^3-18n^2+15n+2}{2}.
\end{eqnarray}
More generally, for any fixed $L$, $\#(n,L)$ is a polynomial of degree
$L$ in $n$.

%
%
Next, consider the set of states which can be obtained as the
non-degenerate ground state~\footnote{We use the ground state for
  concreteness; the argument which follows applies equally to excited
  states.} of an $L$-local Hamiltonian.  This set can be parameterized
by $\#(n,L)$ real parameters.  Since an arbitrary state of $n$ qubits
requires $2\times 2^n-2$ real parameters to specify, provided $\#(n,L)
< 2\times 2^n-2$, we see that there exists a state $\psi$ which cannot
arise as the non-degenerate ground state of any $L$-local Hamiltonian.
Comparing with the bound Eq.~(\ref{eq:number-params}) we see that this
is generically the case except in the case where $L$ approaches $n$,
that is, unless, the number of bodies interacting approaches the
number of bodies in the system.  For large values of $n$ this is an
unphysical situation, and generic quantum states will \emph{not} be
the ground state of a non-degenerate $L$-local Hamiltonian.

%
%
This argument proves the existence of quantum states which are not
eigenstates of any non-degenerate, $L$-body Hamiltonian.  However,
there are many deficiencies with the argument.  First, the argument
only establishes the \emph{existence} of such states, it does not tell
us what they are.  Second, while the argument shows that such a state
cannot be an \emph{exact} eigenstate, it does not provide any
limitation on how close it can be to an eigenstate.  Indeed, phenomena
such as space-filling curves show that a manifold of small dimension
can ``fill up'' a manifold of larger dimension so that every point in
the manifold of larger dimension is arbitrarily close to a point in
the manifold of smaller dimension.  Third, the argument requires the
eigenstates to be non-degenerate.  This deficiency may be partially
remedied by noting that the manifold of states arising as eigenstates
of Hamiltonians with up to $m$-fold degeneracy is at most $m \times
\#(n,L)$-dimensional.  However, as $m$ increases, the bound obtained
by parameter counting becomes weaker and weaker.

%
%
A much stronger argument can be obtained using the theory of quantum
error-correcting codes (QECCs).  We now briefly introduce the relevant
elements of the theory of QECCS, and explain a simple observation
motivating the connection between $L$-local Hamiltonians and QECC
states.  Then, below, we develop a stronger quantitative version of
the argument.

%
%
The idea of quantum error-correction is to encode the state of a small
physical system, such as a qubit, in a larger quantum system, such as
a collection of qubits.  The hope is that the encoded quantum
information will be more robust against the effects of noise than if
it were not encoded.  This hope was realized in schemes proposed by
Shor~\cite{Shor95a} and Steane~\cite{Steane96c}, and since developed
in great detail elsewhere~\footnote{See Chapter~10 of~\cite{Nielsen00a}
  for a review and further references.}

%
%
For example, a code encoding $k$ qubits into $n$ qubits is a
$2^k$-dimensional subspace of the $2^n$-dimensional state space of $n$
qubits.  It is convenient to give the code space a label, $V$.  We say
that the code can correct errors on up to $t$ qubits if the subspaces
$\sigma V$ are all orthogonal to one another, for $\sigma$ of weight
up to $t$.  The idea is that the different $\sigma$ correspond to
different error processes that may occur on the qubits.  Because the
different $\sigma V$ are orthogonal to one another it is possible to
perform a measurement to determine which error occurred, and then
return the system to its original state.  Of course, this does not
address what happens when errors occur that are not simply products of
Pauli matrices on $t$ qubits; perhaps some small random phase rotation
occurs. Remarkably, it turns out that quantum error-correction also
enables us to correct errors which are not products of Pauli matrices;
see Chapter 10 of~\cite{Nielsen00a} for details.

%
%
Strictly speaking, we have described a special type of quantum
error-correcting code, and it is possible to find codes not of this
type.  In particular, for a class of codes known as \emph{degenerate
  codes}, different errors $\sigma$ and $\sigma'$ may have
\emph{identical} effects on the codespace, so $\sigma V$ and $\sigma'
V$ are not orthonormal.  However, for our purposes the non-degenerate
codes we have described above are sufficient.  In particular, there
are many useful bounds on the existence of non-degenerate codes.  We
now describe an example of such a bound.  The bound is the quantum
Gilbert-Varshamov bound, which shows that a code of this type encoding
$k$ qubits into $n$ qubits, and correcting errors on up to $t$ qubits,
exists whenever~\footnote{A different form of the Gilbert-Varshamov bound
  was originally stated in~\cite{Ekert96b,Gottesman97a}.  
  Gottesman~\cite{Gottesman-errata} points out that the earlier
  bound requires
  a slight correction, which we have given here~\cite{Gottesman-GV}.}:
\begin{eqnarray}
  \#(n,2t) < \frac{2^{2n}-1}{2^{n+k}-1}
\end{eqnarray}
In the limit of large $n$ this becomes~\cite{Calderbank97a}
\begin{eqnarray} 
  \frac{k}{n} < 1 - H\left( \frac{2t}{n} \right)-\frac{2t}{n} \log(3),
\end{eqnarray}
where $H(x) \equiv -x \log(x) -(1-x) \log(1-x)$ is the binary entropy,
and all logarithms are taken to base $2$.

%
%
The Gilbert-Varshamov bound applies even when $k = 0$.  Thus there
exists a $1$-dimensional quantum code --- that is, a quantum state,
$\mypsi$ --- such that the states $\sigma \mypsi$ are all orthogonal
to one another.  This is true for $\sigma$ up to weight $t$ for any
$t$ satisfying
\begin{eqnarray} \label{eq:QECC} 
  \#(n,2t) < \frac{2^{2n}-1}{2^n-1}.
\end{eqnarray}
In the large $n$ limit, this becomes $t/n < 0.0946$.  Summarizing, the
quantum Gilbert-Varshamov bound tells us that there exists a quantum
state $\mypsi$ such that the states $\sigma \mypsi$ form an
orthonormal set for $\sigma$ of weight at most $t$, for any $t$
satisfying $\#(n,t) < (2^{2n}-1)/(2^n-1)$.

%
%
Let us return to the problem of Hamiltonians and eigenstates.  Suppose
$\mypsi$ is a state such that $\sigma \mypsi$ form an orthonormal set
for $\sigma$ of weight at most $t$; $\mypsi$ might be a QECC state, as
above.  Expanding $H$ in the form of Eq.~(\ref{eq:op-expansion}), we
see that, provided $L \leq t$, $H\mypsi$ contains terms orthogonal to
$\mypsi$ unless $h_{\sigma} = 0$ for all $\sigma \neq I$.  Thus,
unless $H$ is completely degenerate, $\mypsi$ cannot be an eigenstate
of $H$.  This suggests that QECC states are interesting examples of
states that cannot be eigenstates of local Hamiltonians.  This is
somewhat surprising in light of the fact that QECC states can be
prepared efficiently, i.e., in time polynomial in $n$, on a quantum
computer~\cite{Gottesman97a}.  Indeed, the argument addresses two of
the problems with the parameter counting argument.  Namely, finding a
constructive procedure to find the desired states, $\mypsi$, which can
be done using the methods of quantum error-correction~\footnote{Note
  that efficient, i.e., polynomial in $n$, methods for constructing
  codes which meet the bound in Eq.~(\ref{eq:QECC}) are not known.
  However, a wide range of efficient methods for constructing QECCs
  are known, and even for codes such as those provided by
  Eq.~(\ref{eq:QECC}), finding the codes is an exercise in the theory
  of finite groups that can be solved by enumeration.}, and dealing
with degeneracies in $H$.  However, it leaves the most significant
problem open, namely, proving bounds on how close $\mypsi$ can be to
an eigenstate of $H$.  Remarkably, the answer turns out to be ``not
very'', as we now prove.

%
%
Suppose an $n$-body $L$-local quantum system is described by a
non-trivial Hamiltonian $H$.  We suppose $H$ acts on qubits; the
extension to other systems is straightforward.  Suppose $\myE$ is any
energy eigenstate for the system, with corresponding energy $E$, and
let $H' \equiv H-E I$ be a rescaled Hamiltonian such that $\myE$ has
energy $0$.  Note that $H' = \sum_\sigma h_\sigma' \sigma$, where
$h_I' = h_I-E$, and $h_\sigma' = h_\sigma$ for all other $\sigma$.
Let $\mypsi$ be a state such that $\sigma \psi$ forms an orthonormal
set for $\sigma$ of weight up to $L$, such as a QECC state correcting
errors on $t \geq L$ qubits.  Introducing the operator norm $\|A \|
\equiv \max_{\phi: \| \phi\| = 1} \| A\phi\|$, we have
\begin{eqnarray}
  \| H' (\mypsi - \myE) \| \leq \| H' \| \, \|\mypsi -\myE \|.
\end{eqnarray}
Substituting $H' \myE = 0$, we obtain:
\begin{eqnarray} \label{eq:inter-bound-0}
  \| \mypsi-\myE \| \geq \frac{\| H' \mypsi \|}{\| H' \|}.
\end{eqnarray}
We can assume $\| H' \| \neq 0$, since we have assumed that $H$ is
non-trivial, i.e., it is not a scalar multiple of the identity.  Now,
since the states $\sigma \mypsi$ are orthonormal for all $\sigma$ with
weight at most $L$, we see that:
\begin{eqnarray} \label{eq:inter-bound-1}
\|H' \mypsi\| = \sqrt{\sum_{\sigma} h_\sigma'^2} = \| h' \|_2,
\end{eqnarray}
where $\| \cdot \|_2$ is the Euclidean, or $l_2$, norm for a vector.
Furthermore, by the triangle inequality for norms,
\begin{eqnarray} \label{eq:inter-bound-2}
\| H'\| \leq \sum_\sigma |h_\sigma'| \| \sigma \| = \sum_\sigma |h_\sigma'|
= \| h' \|_1,
\end{eqnarray}
where $\| \cdot \|_1$ denotes the $l_1$ norm of a vector, i.e., the
sum of the absolute value of the components.  Substituting
Eqs.~(\ref{eq:inter-bound-1}) and~(\ref{eq:inter-bound-2}) into
Eq.~(\ref{eq:inter-bound-0}), we obtain
\begin{eqnarray} \label{eq:inter-bound-3}
  \| \mypsi-\myE \| \geq \frac{\| h' \|_2}{\| h' \|_1}.
\end{eqnarray}
The Cauchy-Schwartz inequality tells us that $\|h'\|_1 \leq
\sqrt{\#(n,L)} \| h'\|_2$, where $\#(n,L)$ is the dimension of the
vector $h'$.  Thus we have the general bound
\begin{eqnarray} \label{eq:bound-1}
  \| \mypsi-\myE \| \geq \frac{1}{\sqrt{\#(n,L)}}.
\end{eqnarray}
Eq.~(\ref{eq:bound-1}) provides a constant lower bound on the distance
of $\mypsi$ from any energy eigenstate $\myE$ of $H$, completely
independent of any details about $H$, other than the fact that it is a
non-trivial, $L$-local Hamiltonian, acting on $n$ qubits.

A stronger bound than Eq.~(\ref{eq:bound-1}) can be obtained from
Eq.~(\ref{eq:inter-bound-3}).  To obtain such a bound we need to
remove the dependence of the right-hand-side of
Eq.~(\ref{eq:inter-bound-3}) on the (unknown) parameter $E$.  A
straightforward calculus argument shows that
\begin{eqnarray} 
  \frac{\| h' \|_2}{\| h' \|_1} \geq \frac{1}
    {\sqrt{ 1+ \frac{\left(\sum_{\sigma \neq I} |h_\sigma|\right)^2}
      {\sum_{\sigma \neq I} h_\sigma^2}}},
\end{eqnarray}
and thus
\begin{eqnarray} \label{eq:bound-2}
  \| \mypsi-\myE \| \geq \frac{1}
    {\sqrt{ 1+ \frac{\left(\sum_{\sigma \neq I} |h_\sigma|\right)^2}
      {\sum_{\sigma \neq I} h_\sigma^2}}},
\end{eqnarray}
Note that Eq.~(\ref{eq:bound-1}) can be recovered from
Eq.~(\ref{eq:bound-2}), using a Cauchy-Schwartz argument similar to
that employed above.  

%
%
These results, Eqs.~(\ref{eq:bound-1}) and~(\ref{eq:bound-2}), carry
over directly to \emph{qudit} systems, provided the operator basis
$\sigma$ we expand in is unitary.  The only differences are that (a)
the coefficients $h_{\sigma}$ in Eq.(\ref{eq:bound-2}) may be complex,
and thus it is necessary to work with their modulus, rather than their
actual value; and (b) the value of $\#(n,L)$ in Eq.~(\ref{eq:bound-1})
is somewhat larger for qudit systems.  Combining these results also
with Eq.~(\ref{eq:number-params}), we may summarize these results as a
theorem:

\textbf{Theorem:} Let $H$ be a non-trivial $L$-local Hamiltonian
acting on $n$ qudits.  Let $\mypsi$ be a state such that the states
$\sigma \mypsi$ are orthonormal for all $\sigma$ of weight up to $L$.
(For example, $\mypsi$ might be a QECC correcting errors on up to $L$
qubits.)  Then the following chain of inequalities holds:
\begin{eqnarray}
  \| \mypsi-\myE \| & \geq & \frac{1}
    {\sqrt{ 1+ \frac{\left(\sum_{\sigma \neq I} |h_\sigma|\right)^2}
      {\sum_{\sigma \neq I} |h_\sigma|^2}}} \\
  & \geq & \frac{1}{\sqrt{\#(n,L)}} \\
  & \geq & \left[(L+1) {n \choose L} (d^2-1)^L\right]^{-1/2}. 
\end{eqnarray}

%
%
It is interesting to contrast our results with the theory of naturally
fault-tolerant quantum systems proposed by Kitaev~\cite{Kitaev97c},
and since developed by many researchers.  Such systems possess a
natural resilience to quantum noise processes due to their underlying
physics, rather than requiring complex external control.  This
resilience makes them especially good candidates for quantum
information processing.  A feature of many naturally fault-tolerant
systems is that the ground state of the system Hamiltonian is a
quantum error-correcting code, and thus the system has the desirable
property that at low temperatures it naturally sits in states of the
code.  Our results show that unless the code is degenerate, getting
codes requires extremely non-local Hamiltonians that are implausible
on physical grounds.  Thus, the degeneracy of the quantum codes
appearing in proposals for naturally fault-tolerant quantum systems is
not a fluke, but rather an essential feature necessary for the system
to be resilient to multiple errors.

%
%
To conclude, we have found interesting examples of quantum states far
from the eigenstates of any non-trivial $L$-local Hamiltonian.
Surprisingly, the states we construct can still be prepared
efficiently on a quantum computer.  Our construction has implications
for the physics of locally interacting many-body systems, and for the
theory of naturally fault-tolerant systems for quantum information
processing.

\acknowledgments 

Thanks to Dave Bacon, Patrick Hayden, Alexei Kitaev, John Preskill,
and Ben Schumacher for enjoyable and enlightening discussions.  We
thank Daniel Gottesman for helpful correspondence on the
Gilbert-Varshamov bound, and for permission to use the corrected form
of the bound.  HLH and MAN enjoyed the hospitality of the Institute
for Quantum Information at the California Institute of Technology,
where part of this work was completed.


\begin{thebibliography}{18}
\expandafter\ifx\csname natexlab\endcsname\relax\def\natexlab#1{#1}\fi
\expandafter\ifx\csname bibnamefont\endcsname\relax
  \def\bibnamefont#1{#1}\fi
\expandafter\ifx\csname bibfnamefont\endcsname\relax
  \def\bibfnamefont#1{#1}\fi
\expandafter\ifx\csname citenamefont\endcsname\relax
  \def\citenamefont#1{#1}\fi
\expandafter\ifx\csname url\endcsname\relax
  \def\url#1{\texttt{#1}}\fi
\expandafter\ifx\csname urlprefix\endcsname\relax\def\urlprefix{URL }\fi
\providecommand{\bibinfo}[2]{#2}
\providecommand{\eprint}[2][]{\url{#2}}

\bibitem[{\citenamefont{Nielsen}(1998)}]{Nielsen98d}
\bibinfo{author}{\bibfnamefont{M.~A.} \bibnamefont{Nielsen}}, Ph.D. thesis,
  \bibinfo{school}{University of New Mexico} (\bibinfo{year}{1998}),
  \bibinfo{note}{{arXiv}:quant-ph/0011036}.

\bibitem[{\citenamefont{Nielsen}(2002)}]{Nielsen02e}
\bibinfo{author}{\bibfnamefont{M.~A.} \bibnamefont{Nielsen}},
  \bibinfo{journal}{Sci. Am.} \textbf{\bibinfo{volume}{287}},
  \bibinfo{pages}{66} (\bibinfo{year}{2002}).

\bibitem[{\citenamefont{Preskill}(2000)}]{Preskill00a}
\bibinfo{author}{\bibfnamefont{J.}~\bibnamefont{Preskill}},
  \bibinfo{journal}{J. Mod. Opt.} \textbf{\bibinfo{volume}{47}},
  \bibinfo{pages}{127} (\bibinfo{year}{2000}),
  \bibinfo{note}{{arXiv}:quant-ph/9904022}.

\bibitem[{\citenamefont{Osborne}(2002)}]{Osborne02b}
\bibinfo{author}{\bibfnamefont{T.~J.} \bibnamefont{Osborne}}, Ph.D. thesis,
  \bibinfo{school}{The University of Queensland} (\bibinfo{year}{2002}).

\bibitem[{\citenamefont{Shor}(1995)}]{Shor95a}
\bibinfo{author}{\bibfnamefont{P.~W.} \bibnamefont{Shor}},
  \bibinfo{journal}{Phys. Rev. A} \textbf{\bibinfo{volume}{52}},
  \bibinfo{pages}{2493} (\bibinfo{year}{1995}).

\bibitem[{\citenamefont{Steane}(1996)}]{Steane96c}
\bibinfo{author}{\bibfnamefont{A.~M.} \bibnamefont{Steane}},
  \bibinfo{journal}{Proc. Roy. Soc. Lond. A} \textbf{\bibinfo{volume}{452}},
  \bibinfo{pages}{2551} (\bibinfo{year}{1996}).

\bibitem[{\citenamefont{Nielsen and Chuang}(2000)}]{Nielsen00a}
\bibinfo{author}{\bibfnamefont{M.~A.} \bibnamefont{Nielsen}} \bibnamefont{and}
  \bibinfo{author}{\bibfnamefont{I.~L.} \bibnamefont{Chuang}},
  \emph{\bibinfo{title}{Quantum computation and quantum information}}
  (\bibinfo{publisher}{Cambridge University Press},
  \bibinfo{address}{Cambridge}, \bibinfo{year}{2000}).

\bibitem[{\citenamefont{Calderbank et~al.}(1997)\citenamefont{Calderbank,
  Rains, Shor, and Sloane}}]{Calderbank97a}
\bibinfo{author}{\bibfnamefont{A.~R.} \bibnamefont{Calderbank}},
  \bibinfo{author}{\bibfnamefont{E.~M.} \bibnamefont{Rains}},
  \bibinfo{author}{\bibfnamefont{P.~W.} \bibnamefont{Shor}}, \bibnamefont{and}
  \bibinfo{author}{\bibfnamefont{N.~J.~A.} \bibnamefont{Sloane}},
  \bibinfo{journal}{Phys. Rev. Lett.} \textbf{\bibinfo{volume}{78}},
  \bibinfo{pages}{405} (\bibinfo{year}{1997}).

\bibitem[{\citenamefont{Gottesman}(1997)}]{Gottesman97a}
\bibinfo{author}{\bibfnamefont{D.}~\bibnamefont{Gottesman}}, Ph.D. thesis,
  \bibinfo{school}{California Institute of Technology},
  \bibinfo{address}{Pasadena, CA} (\bibinfo{year}{1997}),
  \bibinfo{note}{{arXiv}:quant-ph/9705052}.

\bibitem[{\citenamefont{Kitaev}(1997)}]{Kitaev97c}
\bibinfo{author}{\bibfnamefont{A.~Y.} \bibnamefont{Kitaev}},
  \bibinfo{journal}{\mbox{arXiv}:quant-ph/9707021}  (\bibinfo{year}{1997}).

\bibitem[{\citenamefont{Osborne and Nielsen}(2002)}]{Osborne02a}
\bibinfo{author}{\bibfnamefont{T.~J.} \bibnamefont{Osborne}} \bibnamefont{and}
  \bibinfo{author}{\bibfnamefont{M.~A.} \bibnamefont{Nielsen}},
  \bibinfo{journal}{Phys. Rev. A} \textbf{\bibinfo{volume}{66}},
  \bibinfo{pages}{032110} (\bibinfo{year}{2002}),
  \bibinfo{note}{{arXiv}:quant-ph/0202162}.

\bibitem[{\citenamefont{Osterloh et~al.}(2002)\citenamefont{Osterloh, Amico,
  Falci, and Fazio}}]{Osterloh02a}
\bibinfo{author}{\bibfnamefont{A.}~\bibnamefont{Osterloh}},
  \bibinfo{author}{\bibfnamefont{L.}~\bibnamefont{Amico}},
  \bibinfo{author}{\bibfnamefont{G.}~\bibnamefont{Falci}}, \bibnamefont{and}
  \bibinfo{author}{\bibfnamefont{R.}~\bibnamefont{Fazio}},
  \bibinfo{journal}{Nature} \textbf{\bibinfo{volume}{416}},
  \bibinfo{pages}{608} (\bibinfo{year}{2002}),
  \bibinfo{note}{{arXiv}:quant-ph/0202029}.

\bibitem[{\citenamefont{Gunlycke et~al.}(2001)\citenamefont{Gunlycke, Bose,
  Kendon, and Vedral}}]{Gunlycke01a}
\bibinfo{author}{\bibfnamefont{D.}~\bibnamefont{Gunlycke}},
  \bibinfo{author}{\bibfnamefont{S.}~\bibnamefont{Bose}},
  \bibinfo{author}{\bibfnamefont{V.~M.} \bibnamefont{Kendon}},
  \bibnamefont{and} \bibinfo{author}{\bibfnamefont{V.}~\bibnamefont{Vedral}},
  \bibinfo{journal}{Phys. Rev. A} \textbf{\bibinfo{volume}{64}},
  \bibinfo{pages}{042302} (\bibinfo{year}{2001}),
  \bibinfo{note}{{arXiv}:quant-ph/0102137}.

\bibitem[{\citenamefont{Vidal et~al.}(2002)\citenamefont{Vidal, Latorre, Rico,
  and Kitaev}}]{Vidal02b}
\bibinfo{author}{\bibfnamefont{G.}~\bibnamefont{Vidal}},
  \bibinfo{author}{\bibfnamefont{J.~I.} \bibnamefont{Latorre}},
  \bibinfo{author}{\bibfnamefont{E.}~\bibnamefont{Rico}}, \bibnamefont{and}
  \bibinfo{author}{\bibfnamefont{A.}~\bibnamefont{Kitaev}},
  \bibinfo{journal}{{arXiv}:quant-ph/0211074}  (\bibinfo{year}{2002}).

\bibitem[{\citenamefont{Preskill}(1998)}]{Preskill98c}
\bibinfo{author}{\bibfnamefont{J.}~\bibnamefont{Preskill}},
  \emph{\bibinfo{title}{Physics 229: Advanced mathematical methods of physics
  --- Quantum computation and information}} (\bibinfo{publisher}{California
  Institute of Technology}, \bibinfo{address}{Pasadena, CA},
  \bibinfo{year}{1998}),
  \bibinfo{note}{http://www.theory.caltech.edu/people/preskill/ph229/}.

\bibitem[{\citenamefont{Ekert and Macchiavello}(1996)}]{Ekert96b}
\bibinfo{author}{\bibfnamefont{A.}~\bibnamefont{Ekert}} \bibnamefont{and}
  \bibinfo{author}{\bibfnamefont{C.}~\bibnamefont{Macchiavello}},
  \bibinfo{journal}{Phys. Rev. Lett.} \textbf{\bibinfo{volume}{77}},
  \bibinfo{pages}{2585} (\bibinfo{year}{1996}),
  \bibinfo{note}{{arXiv}:quant-ph/9602022}.

\bibitem[{\citenamefont{Gottesman}()}]{Gottesman-errata}
\bibinfo{author}{\bibfnamefont{D.}~\bibnamefont{Gottesman}},
  \emph{\bibinfo{title}{Errata for~\cite{Gottesman97a}}}.

\bibitem[{\citenamefont{Gottesman}(2003)}]{Gottesman-GV}
\bibinfo{author}{\bibfnamefont{D.}~\bibnamefont{Gottesman}}
  (\bibinfo{year}{2003}), \bibinfo{note}{private communication}.

\end{thebibliography}

\end{document}